\documentclass[pre,twocolumn,aps,superscriptaddress,showpacs,floatfix]{revtex4}
\usepackage{graphicx}
\usepackage{dcolumn}
\usepackage{bm}
\usepackage{amsmath}
\begin{document}
\title{Temperature-resonant cyclotron spectra in confined geometries}

\author{A. Pototsky}
\affiliation{Department of Mathematics, University of Cape Town,
Rondebosch 7701, South Africa}
\author{P. H{\"a}nggi}
\affiliation{Institut f{\"u}r Physik, Universit{\"a}t Augsburg,
D-86135 Augsburg, Germany}
\author{F. Marchesoni}
\affiliation{Dipartimento di Fisica, Universit\`a di Camerino,
I-62032 Camerino, Italy}
\author{S. Savel'ev}
\affiliation{Department of Physics, Loughborough University,
Loughborough LE11 3TU, United Kingdom}

\begin{abstract}
We consider a two-dimensional gas of colliding charged particles
confined to finite size containers of various geometries and subjected
to a uniform orthogonal magnetic field. The gas spectral densities are
characterized by a broad peak at the cyclotron frequency. Unlike for
infinitely extended gases, where the amplitude of the cyclotron peak
grows linearly with temperature, here confinement causes such a peak
to go through a maximum for an optimal temperature. In view of the
fluctuation-dissipation theorem, the reported resonance effect has a direct counterpart in the electric susceptibility of the confined magnetized gas.
\end{abstract}

\maketitle
\section{Introduction}
Electronic fluctuations are of a great importance in plasma physics,
due to their relevance in charge and energy transport \cite{gentle_plasma}
and to the well-established connection between fluctuation spectra and
electronic susceptibility \cite{sosenko94}. As the temperature of the
system is increased, the power of thermal fluctuations contained in any
narrow frequency interval is also expected to increase. However, such a
straightforward temperature dependence has been observed in systems where
the electron dynamics is incoherent, as a result of the electron interactions
with other electrons, positively charged ions, and impurities. If, however,
the electron dynamics also contains a coherent component such as, for instance,
rotation with cyclotron frequency in the presence of a magnetic field, then the
temperature dependence of the fluctuation power spectra may develop nontrivial
resonant behaviors. For instance, it has been demonstrated \cite{li94_resonance}
that a magnetized plasma operated at the limit-cycle fixed-point bifurcation
point and driven by a tunable external white noise undergoes, stochastic
resonance \cite{SR}. On the other hand, the observation that in constrained
geometries the matching of thermal scales and characteristic system length
can produce detectable resonant effects has been previously reported \cite{burada,geoSR}.

In this paper we show that a simpler instance of temperature controlled
resonance can naturally occur in a confined magnetized electron gas, due to the
matching of two {\it lengths}, the electron intrinsic thermal length,
or gyroradius, and the finite system size. The effect investigated here
should not be mistaken for a manifestation of the well know electron cyclotron resonance,
which results, instead, from the matching of two {\it frequencies}, the cyclotron frequency
of an electron moving in a uniform magnetic field and the pump frequency of a perpendicular ac electric field \cite{cyclo res}.
The dynamics of a magnetoplasma electron can be reduced to the two-dimensional (2D) Brownian motion of a charged particle subjected to a uniform magnetic field.
In Sec. \ref{infinite} we analyze the power spectral density of a
magnetized Brownian particle moving in an unconstrained planar geometry. In particular,
we notice that the amplitude of the cyclotron peak grows linearly with temperature.
At variance with this remark, in Sec. \ref{finite} our numerical simulations show
that in constrained geometries the cyclotron peak goes through a maximum for an
optimal temperature (Sec. \ref{effect}), which, in turn, is determined by the
matching of system size and (temperature dependent) average cyclotron
radius (Sec. \ref{analytic}). In Sec. \ref{geometries} we also show that
the observed resonant temperature dependence of the cyclotron peak is robust
with respect to variations of the boundary conditions and the geometry of the
system. Finally, in Sec. \ref{conclusions} we discuss possible applications of
this effect to confined systems of magnetocharges in biological and artificial structures.

\begin{figure}[btp]
\centering
\includegraphics[width=0.40\textwidth]{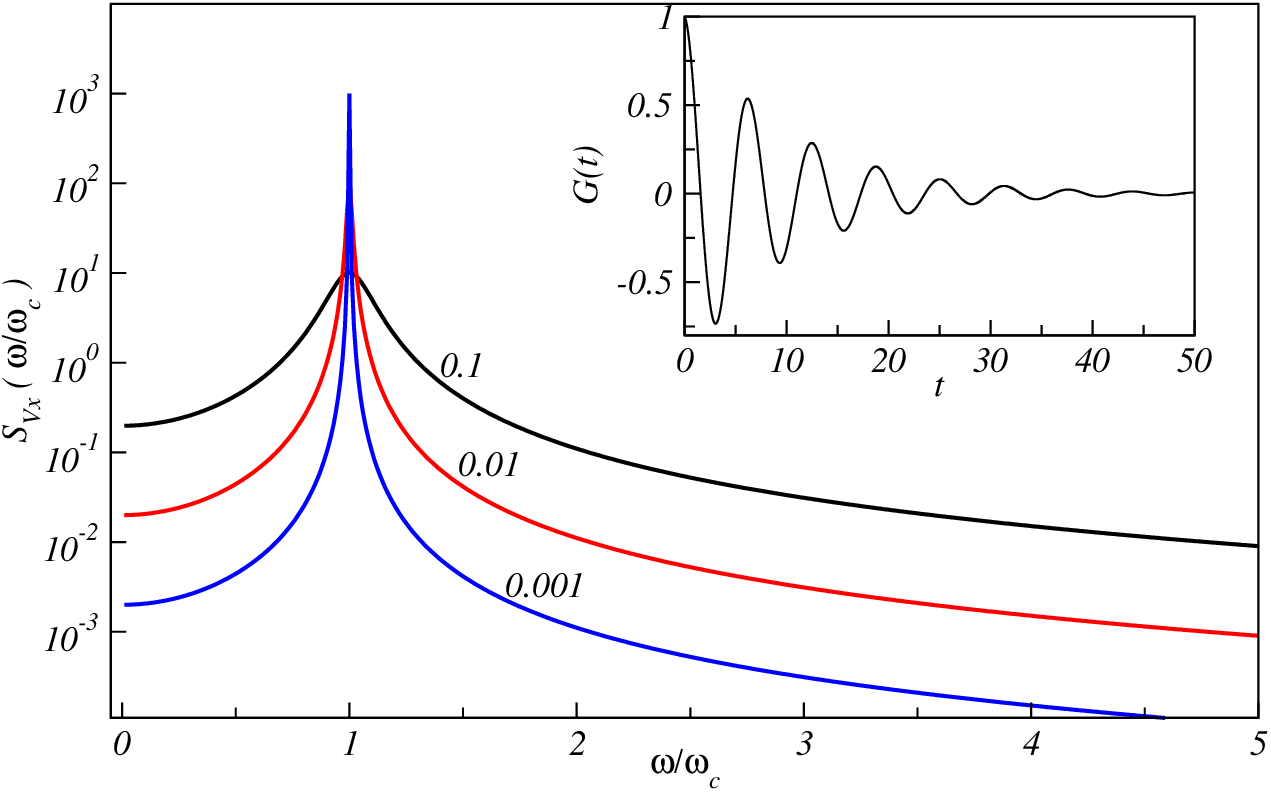}
\caption{(Color online)  Power spectral density $S_{v_x}(\omega) = S_{v_y}(\omega)$,
Eq.\,(\ref{PS}), for $kT/m=1$, $\omega_c=1$ and different $\gamma$ (reported in the legend).
Inset shows the velocity correlation function of Eq.\,(\ref{Corr}), $G(t)$,  for $\gamma=0.1$. \label{F1}}
\end{figure}

\section{Unbounded electron gas}
\label{infinite}

In an equilibrium neutral plasma electrons of charge $q$ and mass $m$
oscillate with characteristic angular frequency \cite{chen_plasma} $\omega_p = [n_0
q^2/(m\varepsilon_0)]^{1/2}$ (in S.I. notation), where $n_0$ is the
average electron density and $\epsilon_0$
the vacuum permittivity. In the following we restrict ourselves to weakly magnetized plasmas to ensure that the cyclotron frequency associated with $B_0$, $\omega_c$, is much smaller than $\omega_p$, i.e., $\omega_c=qB_0/m \ll \omega_p$. This allows us to reduce the dynamics of a magnetoplasma electron to the 2D Brownian motion of a charged particle subjected to a uniform magnetic field. This is a
longstanding problem in plasma and astroparticle physics \cite{magLE,Garb,additional_refs}. Here we limit ourselves to introduce the results relevant to the discussion of our simulation data.

The corresponding Langevin equation reads
\begin{eqnarray}
\label{lang}
&&\vec{v}=\dot{\vec{r}}, \nonumber \\
&&\dot{\vec{v}} = \frac{q}{m} (\vec{v}\times \vec{B}_0) - \gamma \vec{v} +\sqrt{2\gamma\frac{kT}{m}} ~\vec{\xi}(t),
\end{eqnarray}
%
where the vector $\vec{\xi}(t) = (\xi_x(t),\xi_y(t))$ represents two
independent Gaussian white noises with $\langle \xi_i(t)\rangle=0$ and $\langle \xi_i(t) \xi_i(0)\rangle = \delta (t)$ for $i=x,y$.

For numerical purposes, it is convenient to rescale both time, $t
\to \omega_c t$, and space, $r \to r/\lambda$. We recall that
$\omega_c$ is the $B_0$-dependent cyclotron angular frequency and
$\lambda$ is the $T$-dependent electron gyroradius
$\lambda=\sqrt{kT/(m\omega_c^2)}$ \cite{chen_plasma}. In
dimensionless units Eq.\,(\ref{lang}) reads
\begin{eqnarray}
\label{lang2}
\dot{\vec{r}} &=& \vec{v}, \nonumber\\
\dot{\vec{v}} &=&  \vec{v}\times \vec{b}_0 - g \vec{v} +\sqrt{2g} \vec{\xi}(t),
\end{eqnarray}
where $\vec{b}_0$ is a unity vector parallel to $\vec{B}_0$. Note that,
in the absence of boundaries, the only free parameter in the dimensionless
Langevin equation (\ref{lang2}) is the scaled damping constant $g=\gamma/\omega_c$.
In the foregoing sections all results will be given in dimensional units for reader's convenience.

For the linear and unconstrained dynamics of Eq.\,(\ref{lang2}) the
p.s.d. $S(\omega) = \langle | \hat{\vec{r}}(\omega)|^2\rangle$, with
$\hat{\vec{r}}(\omega)$ standing for the Fourier transform of
$\vec{r}(t)$, can be computed analytically by means of standard
harmonic analysis \cite{Gard,HT82}. Taking advantage of the fact
that $\xi_x(t)$ and $\xi_y(t)$ are uncorrelated white Gaussian
noises, we obtain
\begin{eqnarray}
\label{PS}
S_x(\omega) &=& S_y(\omega) = \frac{S_{v_x}}{\omega^2},  \\
S_{v_x}(\omega) &=&   \frac{2kT}{m}\frac{\gamma(\omega^2+\gamma^2)
(\omega^2+\gamma^2+\omega_c^2)}{\gamma^2\left( \omega^2 + \gamma^2 +
\omega_c^2\right)^2 + \omega^2 \left( \omega^2+\gamma^2-\omega_c^2\right)^2}, \nonumber
\end{eqnarray}
where $S_i$ and $S_{v_i}$ denote the p.s.d. of the $i=x,y$
components of the 2D vectors $\vec{r}$ and $\vec{v}$, respectively.
Note that at resonance $\omega=\omega_c$, the peak of the transverse
velocity is $S_{v_x}(\omega_c) = v^2_{\rm th}/(2\gamma)$, with
$v_{\rm th}^2=\langle {\vec{v}}^{\;2} \rangle=2kT/m$. In view of the
discussion below, we recall here that the stationary autocorrelation
function of the velocity, $G(t-t^\prime) = \langle v_x (t)
v_x(t^\prime)\rangle = \langle v_y (t) v_y(t^\prime)\rangle$, solely
depends on the difference $t-t^\prime$ and is related to the inverse
Fourier transform of the p.s.d. $S_{v_x}(\omega)$ through the
Wiener-Khinchin theorem \cite{Garb,additional_refs,Gard,HT82},
\begin{eqnarray}
\label{Corr}
G(t-t^\prime) &=& \langle v_x(t)v_x(t^\prime)\rangle = \frac{1}{2\pi}
\int_{-\infty}^{\infty}S_{v_x}(\omega)e^{i\omega (t-t^\prime)}\,d\omega \nonumber \\
&=& \frac{kT}{m}e^{-\gamma\mid t-t^\prime\mid}\cos{[\omega_c(t-t^\prime )]}.
\end{eqnarray}
For an unbounded planar electron gas, the stationary distribution density of the velocity is
Maxwellian \cite{magLE,Garb,additional_refs} and does not depend on either the damping constant,
$\gamma$, or the magnetic field, $B_0$,
\begin{eqnarray}
\label{Max}
f(\vec{v}) = \frac{m}{2\pi kT}\exp{\left( -\frac{m\vec{v}\,^2}{2kT}\right)}.
\end{eqnarray}
The autocorrelation function of the electron coordinates,
$\langle \vec{r}(t) \vec{r}(t^\prime) \rangle$, diverges as a function of $t$
and $t^\prime$ because the free motion of electrons on the plane is unbounded.

To this regard it should be noticed that the diffusion of a Brownian charge carrier on a
plane perpendicular to a constant magnetic field is normal, that is, for asymptotically
large $t$, $\langle \vec{r}\,^2\rangle=4D_Bt$ with
\begin{equation}
\label{DB}
D_B=\frac{kT}{m\gamma}~\frac{\gamma^2}{\gamma^2+\omega_c^2}.
\end{equation}
This means that for $B_0>0$ the particle diffusivity gets suppressed, as $D_B$ is smaller
than the free diffusion coefficient $D_0=kT/(m\gamma)$.

The typical velocity p.s.d., $S_{v_x}(\omega)$, and an example of
velocity autocorrelation function, $G(t)$, are depicted in
Fig.\,\ref{F1} for different values of the damping constant
$\gamma$. The peak at the cyclotron frequency $\omega_c$ broadens as
$\gamma$ is increased. In the remaining sections of this paper we
choose $\gamma$ to be much smaller than $\omega_c$, i.e. $\gamma \ll
\omega_c$: This ensures that the cyclotron peak is well pronounced
or, equivalently, that electrons with any given velocity $v$ tend to
perform many cyclotron orbits of radius $r_c=v/\omega_c$, before
being perturbed by the combined action of noise and friction.
Accordingly, in the underdamped limit the diffusion coefficient
$D_B$ tends to $kT\gamma/(m\omega_c^2)=\lambda^2\gamma$.

\section{Finite systems}
\label{finite}

Our goal now is to compute the power spectral densities (p.s.d.), $S(\omega)$,
of a confined 2D gas of electrons with finite temperature and for different
geometries and boundary conditions. It should be noticed that $S(\omega)$
encodes important information about the electromagnetic transmission
properties of the electron gas. In fact, $S(\omega)$ can be directly
linked to the imaginary part of the gas electric susceptibility, $\kappa(\omega)$,
via the fluctuation-dissipation relation, $S(\omega) \sim kT {\rm Im}[\kappa(\omega)]$ \cite{sosenko94}.

We start with the simplest case of an electron gas trapped in a strip delimited by two
walls parallel to the $y$ axis and a fixed distance $d$ apart. Electrons are assumed to
be reflected elastically by the strip boundaries, which leads to a zero net transverse flux in the $x$ direction.

We numerically integrated the dimensionless Langevin equation
(\ref{lang2}) for $g$ in the range $10^{-3}\div 10^{-2}$ and then
restored dimensional units with $\omega_c=1$ and fixed system size.
With respect to the time unit, this corresponds to reporting
$\omega$ in units of $\omega_c$, with no severe restriction on the
actual value of $B_0$, but for the condition that $\omega_c \ll
\omega_p$ (see Secs. \ref{infinite} and \ref{conclusions} for more
details). With respect to the space units, setting the width of the
strip to a given value, $d$, makes the electron p.s.d. depend
explicitly on the temperature, which had been eliminated from Eq.
(\ref{lang2}) by expressing all lengths in units of $\lambda$. Note
that in our plots $\omega_c=1$, so that the scaled temperature
$kT/m$ boils down to the square of the gyroradius, $\lambda$,
introduced in Sec. \ref{infinite}.

\begin{figure}[btp]
\centering
\includegraphics[width=0.45\textwidth]{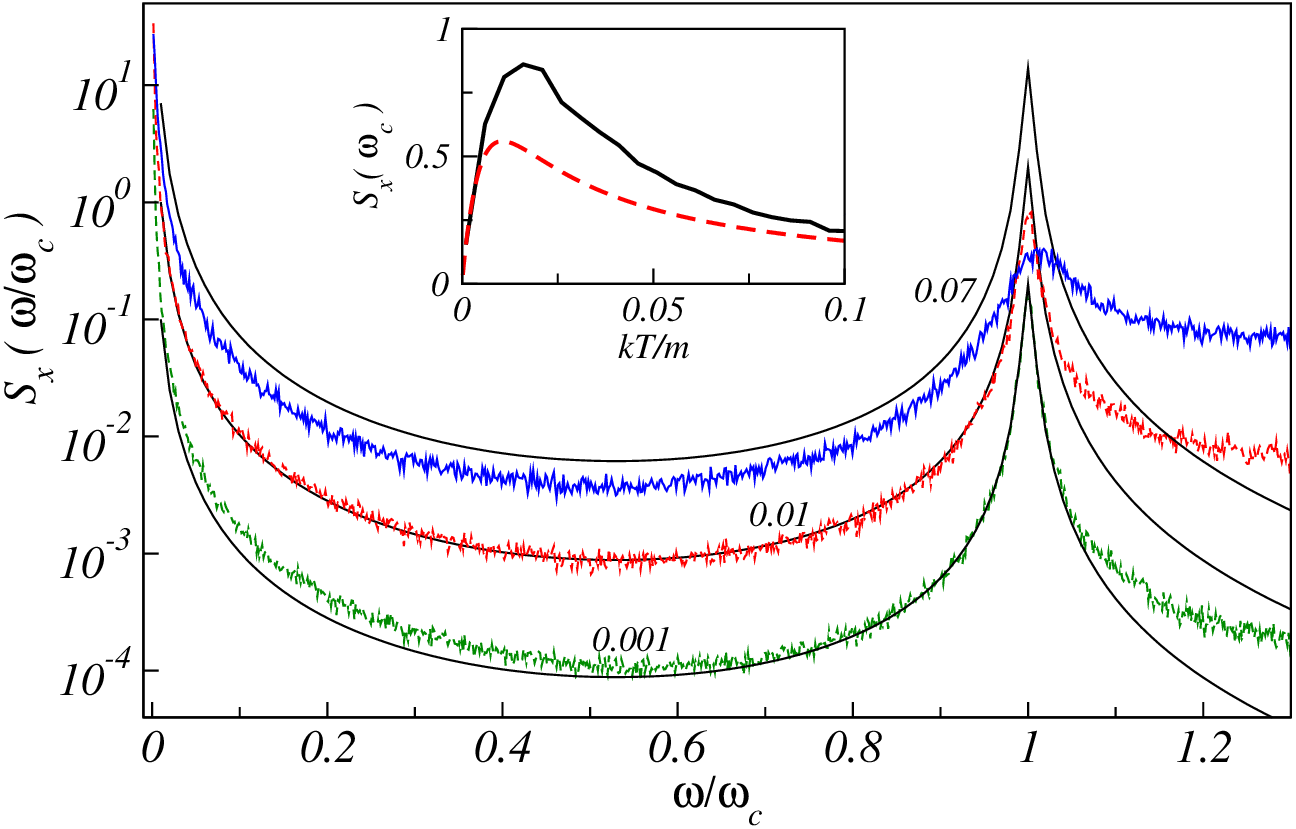}
\caption{(Color online)  Transverse power spectral density, $S_x(\omega)$,
of an electron moving in a strip of width $d=1$ (see sketch in Fig. \ref{F3}),
as computed from Eq.\,(\ref{lang2}) for $\gamma=0.005$, $\omega_c=1$ and the
scaled temperature $kT/m$ reported in the legend. Solid curves represent the
corresponding p.s.d. for an unbounded planar system. Inset: amplitude of the cyclotron peak $S_x(\omega_c)$ as a function of $kT/m$ (solid curve) and its analytical approximation, Eq. (\ref{approx}) (dashed curve). Note that for
$\omega_c=1$ the scaled temperature $kT/m$ coincides with $\lambda^2$. \label{F2}}
\end{figure}

\subsection{Resonant cyclotron peak}
\label{effect}

The typical p.s.d. of the transverse coordinate,  $S_x(\omega)$, are
depicted in the main panel of Fig.\,\ref{F2}. The finely dotted
curves represent $S_x(\omega)$, as computed numerically through
Eq.\,(\ref{lang2}) with $\gamma=0.005$ and ideal reflecting
boundaries located at $x=\pm d/2$ with $d=1$. The solid curves are
the corresponding $S_x(\omega)$ for an unbounded planar electron
gas, as predicted in Eq.\,(\ref{PS}). The height as well as the
width of the cyclotron peak depend on the scaled temperature $kT/m$.

Unlike for the case of an infinite system, the amplitude of the
cyclotron peak, $S_x(\omega_c)$, for the confined electron gas
depends resonantly on the temperature, as shown in the inset of
Fig.\,\ref{F2}: As $T$ is gradually increased, $S_x(\omega_c)$ goes
through a maximum for an optimal temperature, $T_c$, whose
dependence on the confinement geometry is investigated in the
following sections. This behavior may be reminiscent of stochastic
resonance \cite{SR}. However, we anticipate that here the optimal
temperature is defined by the matching of two length scales, rather
than two time scales, as it is the case in ordinary stochastic
resonance \cite{SR,JH91}. Finite volume effects have been reported in the early stochastic resonance literature \cite{phi4}, but in a totally different context.
\subsection{Quantitative interpretation}
\label{analytic}
\begin{figure}[btp]
\centering
\includegraphics[width=0.45\textwidth]{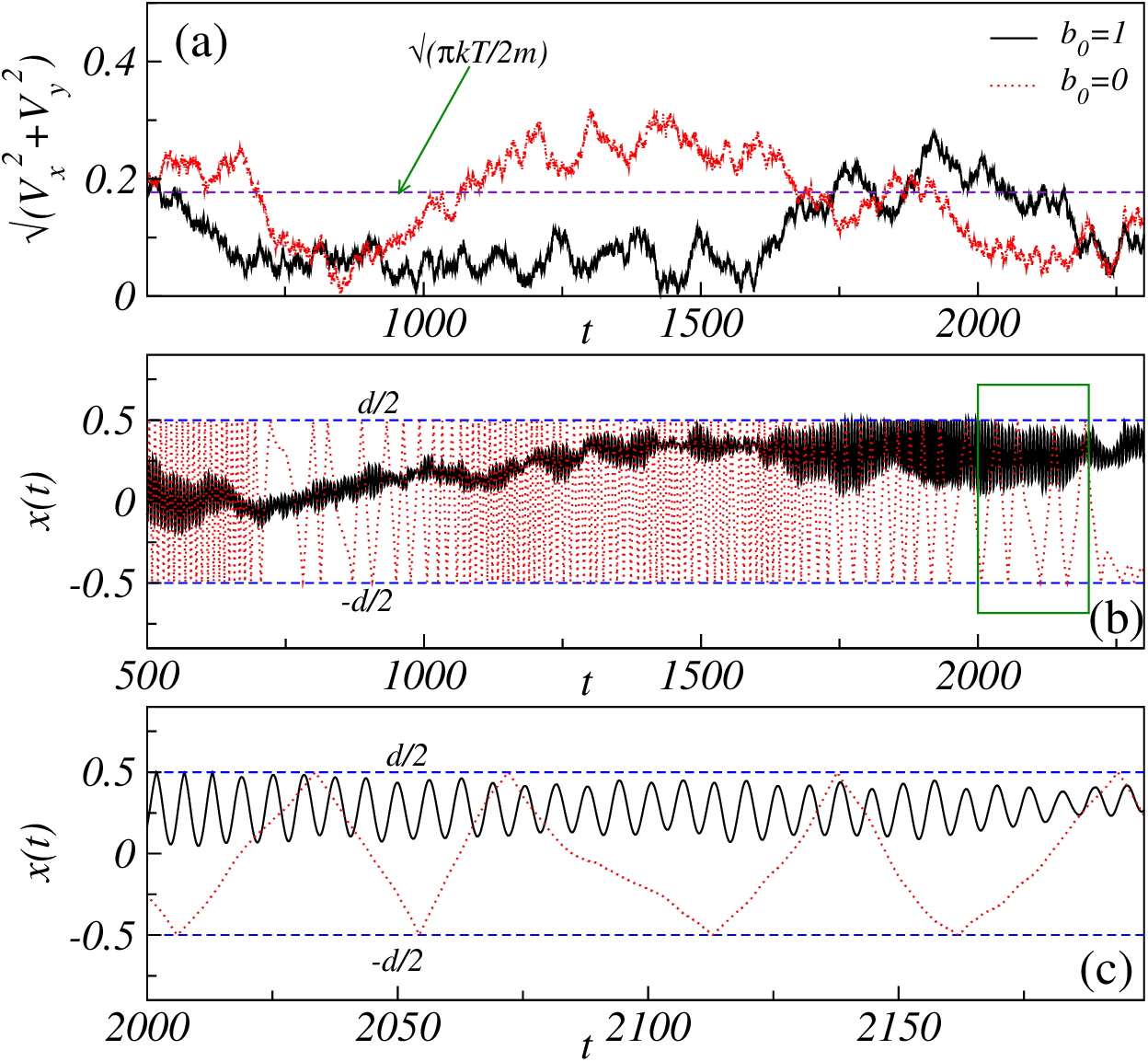}
\caption{(Color online) Trajectory samples of $|\vec{v}|(t)$, (a), and $x(t)$,
(b)-(c), for $kT/m=0.02$, $\gamma=10^{-3}$, $d=1$ and $\omega_c=1$
(black curves) or $B_0=0$ (red curves). In (c) is a blow-up of the
trajectory portion marked by a rectangle in (b) The horizontal dashed line in
(a) represents $\langle |\vec{v}| \rangle$ as computed from Eq. (\ref{vel}). \label{F3}}
\end{figure}

The resonant temperature dependence of the cyclotron peak can be
qualitatively explained as follows. First we recall that, in the
underdamped regime, the diffusion time of a charged Brownian
particle across a strip of width $d$ is strongly suppressed by the
presence of a magnetic field, especially for $\gamma t \ll 1$. From
Eq. (43) of Ref. \cite{Garb}, $\langle \vec{r}\,^2\rangle =
{\cal{O}}(t^3)$; hence, the transverse diffusion time in the strip
can be safely assumed to be much shorter than the cyclotron period,
as illustrated in panels (a) and (b) of Fig. \ref{F3}.

Then, we notice that, as the interaction of the electrons with the walls is elastic,
the equilibrium distribution of their velocity is not affected by the boundary
geometry and is still given by the Maxwell distribution of Eq.\,(\ref{Max}).

Moreover, the cyclotron radius of an electron moving with instantaneous velocity
$v$ is $r_c = v/\omega_c$; in the regime of low damping, $\gamma \ll \omega_c$,
we can neglect the effects of friction on the electron orbits. This means that the
contribution to $S_x$ from a circular orbit of constant radius $r_c$ is proportional
to $r_c^2/2\gamma$ [see discussion following Eq. (\ref{PS})], that is, increases
quadratically with $v$. However, this conclusion applies only to electronic
trajectories with centers located a distance not smaller than $r_c$ away from
the reflecting boundaries.  Indeed, when the electrons come too close to the boundaries,
they repeatedly bounce off the walls, so that their trajectories get distorted
[see Fig. \ref{F3}(c)]. For an equilibrium distribution of the electron velocities,
this surely happens when their orbit diameter is larger than half the strip width, $2r_c>d/2$. 

%
\begin{figure*}[btp]
\centering
\includegraphics[width=0.8\textwidth]{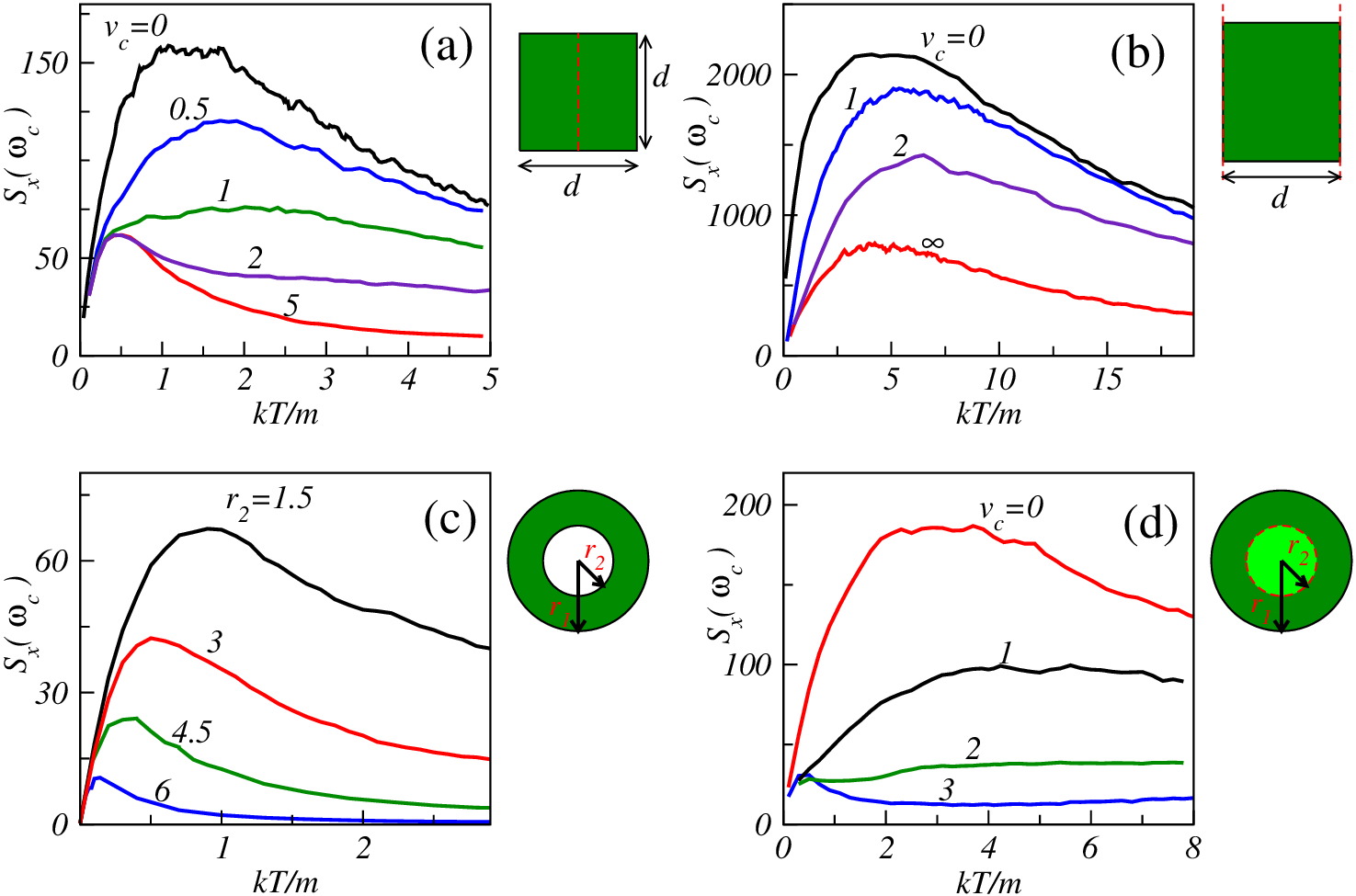}
\caption{(Color online) Amplitude of the cyclotron peak,
$S_x(\omega_c)$, as a function of the scaled temperature $kT/m$ for
the different geometries of Sec. \ref{geometries} as sketched. In
all four panels $\omega_c=1$ and $\gamma=0.00166$ (a), (b) and
$0.003$ (c), (d).  (a) Square box containing a semi-transparent
wall $x=h$ [case (i)]. Simulation parameters are: $d=12$, $h=0$,
and threshold velocity $v_c$ as reported next to the relevant data
sets. (b) [case (ii)] Infinite strip with semi-transparent
walls. Simulation parameters are: $d=18$ and $v_c$ as reported
next to the relevant data sets. (c) Annulus with reflective inner
boundary [case (iii)] with $v_c=\infty$. The outer radius is kept
constant, $r_1=9$, and the inner radius, $r_2$, varied as in the
legend. (d) [case (iv)] Same geometry as in (c) but with constant
inner radius, $r_2=4.5$, and semi-transparent inner wall with
threshold velocity, $v_c$, in the legend. \label{F4}}
\end{figure*}

In view of the arguments above, we assume for simplicity that fast electrons with too
large a cyclotron radius, say, $r_c>d/4$, do not contribute to the cyclotron peak,
whereas only a fraction $1-4r_c/d$ of the slower electrons with $r_c<d/4$ do.
On further noticing that from Eq. (\ref{Max}) the equilibrium distribution of $v$ is
\begin{eqnarray}
\label{vel}\rho(v) =\frac{mv}{kT}\exp{\left( -
\frac{mv^2}{2kT}\right)}\,,
\end{eqnarray}
we obtain the following estimate for the amplitude of the cyclotron peak,
\begin{eqnarray}
\label{approx}
S_x(\omega_c) \simeq \frac{(\bar Td)^2}{32\gamma}\left [2+e^{-\frac{1}
{2\bar T^2}}-3\sqrt{\frac{\pi}{2}}\,\bar T\,{\rm Erf}\left(\frac{1}{\sqrt {2 \bar T^2}}\right)\right],
\end{eqnarray}
where ${\bar T}=4\sqrt{kT/m}/(d\omega_c)$ and Erf($\dots$) denotes the standard error function.
%
Note that this estimate for $S_x(\omega_c)$ systematically underestimates the
corresponding simulation curve as we neglected the residual contribution from
electronic orbits larger, but not too larger, than $d/4$. By numerically evaluating
Eq. (\ref{approx}), one concludes that both the resonance value of the cyclotron peak,
$S_{\rm max}(\omega_c)$, and $T_c$ grow quadratically with $d$, namely,
$kT_c/m \simeq 0.01(\omega_c d)^2$ and $S_{\rm max}(\omega_c)\simeq 2.8\cdot 10^{-3}d^2/\gamma$.

We also stress that the emergence of a resonant cyclotron peak is not conditioned by the elastic boundary assumption. Boundary randomness or fluctuations may, indeed, affect the residual contribution from large electronic orbits with $r_c>d/4$, but not the bulk contribution estimated in Eq. (\ref{approx}).

\section{Dependence on geometry and boundary conditions}
\label{geometries}
Next we numerically compute $S_x(\omega_c)$ as a function of the temperature for
different geometries and boundary conditions, in order to demonstrate the robustness
of the temperature resonance of the cyclotron peak.

We consider here three confining setups for the 2D electron gas:

{(i) \it Box with internal semi-transparent wall.}\\
Consider a $d \times d$ square box, centered at the origin $x=y=0$,
and containing a semi-transparent internal wall, $x=h$ with
$|h|<d/2$, parallel to the $y$ axis. The internal wall acts like a
porous filter letting charges pass through only if the $x$ component
of their velocity, $v_x$, is larger than a certain threshold
velocity $v_c$, i.e. $v_x > v_c$. If $v_x\leq v_c$, the electrons
are elastically reflected back into their half box. We fix $d=12$,
$h=0$ and plot in Fig.\,\ref{F4}(a) the amplitude of the cyclotron
peak as a function of the scaled temperature $kT/m$ for different
values of the threshold. For $v_c=0$ the compartment wall is
transparent, so that the effective width of the box is $d$, whereas
for $v_c>5$ the internal wall acts as an almost perfectly reflective
boundary, thus dividing the square box in two rectangular
compartments of width $d/2$. Despite the different geometries,
confined cyclotron orbits are confirmed to generate a resonant
temperature dependence of the electronic p.s.d., no matter what
$v_c$. Moreover, in agreement with our analytical discussion of the
infinite strip from Sec. \ref{analytic}, the optimal temperature
$T_c$, corresponding to the maxima of the plotted curves, diminishes
with, $ d^2 \longrightarrow d^2/4$,  by a factor $4$ on increasing
$v_c$ from $0$ to $\infty$ and, thus, halving the container width.

{(ii) \it Infinite strip with semi-transparent walls.}\\
Let us consider the infinite strip of Sec. \ref{effect} with the important
difference that now its parallel walls are semi-transparent, as
described in $(i)$. If $v_x \leq v_c$ the electrons are contained in the strip;
if $v_x > v_c$ the electrons exit one wall and reenter through the other one
with periodic boundary conditions. In Fig.\,\ref{F4}(b), the peak amplitude $S(\omega_c)$
is plotted as a function of the temperature for different values of the threshold.
Since the boundaries are periodic for $v_c=0$ and reflective for $v_c = \infty$,
here our approximate estimate for $S(\omega_c)$ from Sec. \ref{analytic}
is expected to work well only as $v_c \to \infty$. In this limit,
Eq. (\ref{approx}) reproduces fairly closely both $T_c$ and  $S_{\rm max}(\omega_c)$.
In the opposite limit of purely periodic boundary conditions, the cyclotron peak at
resonance, $S_{\rm max}(\omega_c)$, gets enhanced, while $T_c$ only weakly depends on $v_c$.

{(iii) \it Annulus with with a reflecting inner wall} and {(iv) \it with a semi-transparent inner wall.}\\
Next let the electrons be trapped in a circle with radius $r_1$,
which represent an ideal reflecting boundary. The inner space is
divided by a second circle of radius $r_2$, with $r_2<r_1$, into two
compartments. The inner circle is concentric with the outer circle
and its circumference works as a semi-transparent wall (case (iv))
with threshold velocity $v_c$ (applied to the radial component of
${\vec v}$). The case  (iii) of an ideal reflecting inner circle,
corresponds to setting the threshold velocity $v_c=\infty$, see
Fig.\,\ref{F4}(c), the gas is confined to an annulus. As one can
anticipate from the discussion in Sec. \ref{analytic}, the maximum
of the cyclotron peak, $S_{\rm max}(\omega_c)$, decreases on
increasing $r_2$, see Fig.\,\ref{F4}(c). Correspondingly, the
optimal temperature, $T_c$, also decreases because the width of the
annulus shrinks.
The dependence of the resonant cyclotron effect on $v_c$ is illustrated in
Fig.\,\ref{F4}(d). The effect is most pronounced in the case of a perfectly
transparent inner circle, $v_c=0$. Indeed, lowering $v_c$ enlarges the surface
accessible to the cyclotron orbits of the confined electrons. As suggested by
Eq. (\ref{approx}), $T_c$ and the maximum of $S(\omega_c)$ grow quadratically
with the effective transverse dimensions of the gas container; for the simulation
parameters reported in Fig.\,\ref{F4}(d), this corresponds to an increase of both
quantities by a factor of about $4$ as $v_c$ raises from $0$ to $\infty$.

\begin{figure}[btp]
\centering
\includegraphics[width=0.35\textwidth]{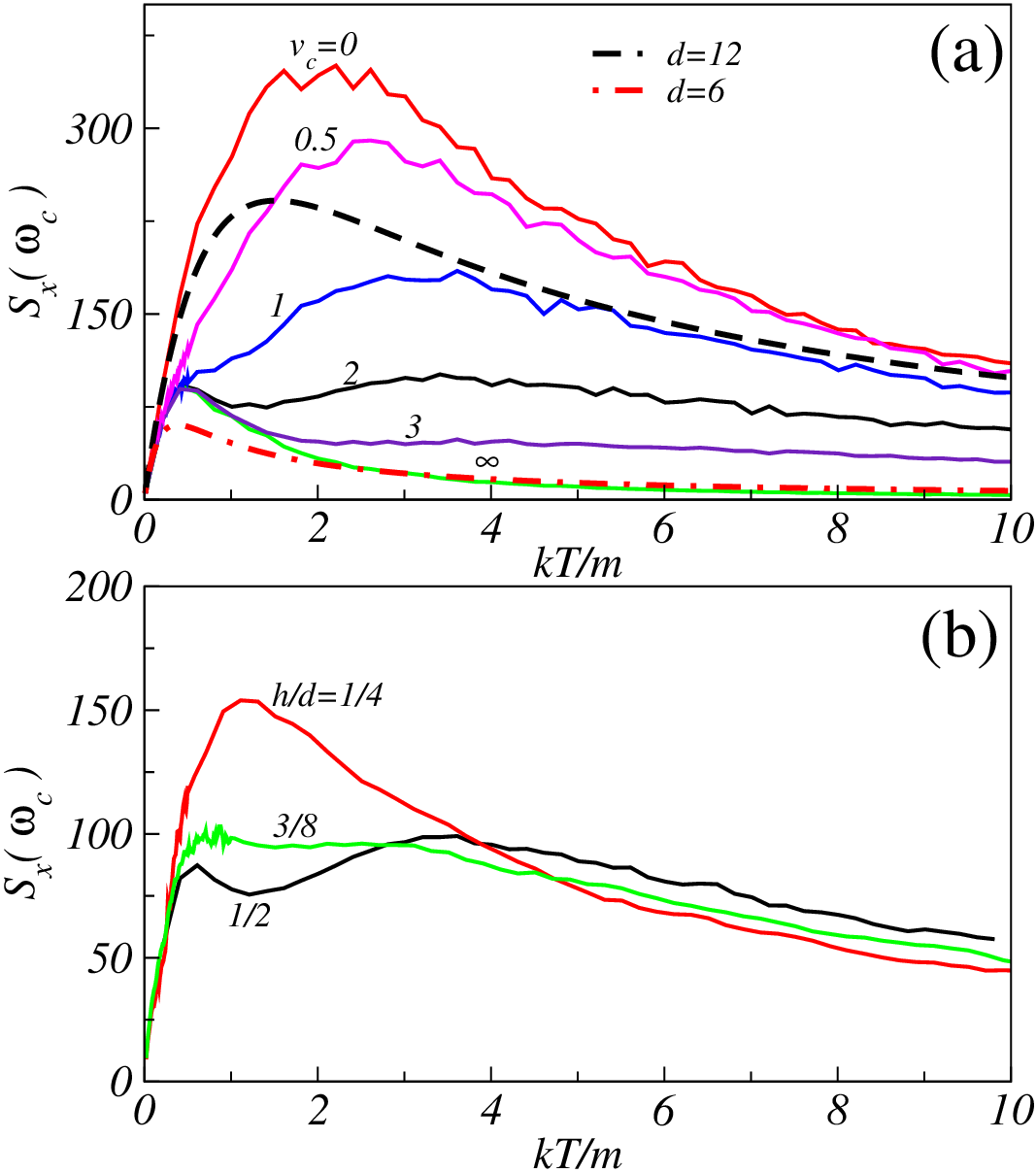}
\caption{(Color online) Amplitude of the cyclotron peak,
$S_x(\omega_c)$, as a function of the scaled temperature $kT/m$
[case (v) of Sec. \ref{geometries}] for $\omega_c=1$ and
$\gamma=0.00166$. (a) Infinite strip of width $d=12$ with a
semi-transparent filter located at $h=0$ and different thresholds
$v_c$ (in the legend). The curves of Eq. (\ref{approx}) for $d=12$ (dashed) and $6$ (dotted-dashed) are reported for a comparison; (b) Geometry as in (a), $v_c=2$ and different
$h$ (in the legend). \label{F5}}
\end{figure}

{(v) \it Infinite strip with internal semi-transparent wall.}\\
Finally, we show that by appropriately choosing the geometry of the
system, the cyclotron peak can go through two maxima as a function
of temperature. Such a double resonance was found by inserting an
internal wall, $x=h$ with $|h|<d/2$, of tunable threshold $v_c$ in
the infinite strip of Sec. \ref{effect}. Our simulation results for
a symmetric geometry with $d=12$, $h=0$ and different $v_c$ are
displayed in Fig.\, \ref{F5}(a). The limiting regimes, $v_c=0$ and
$v_c \to \infty$ are well reproduced by our approximate formula in
Eq. (\ref{approx}). In the intermediate regimes, say at $v_c=2$, the
$S(\omega_c)$ features two maxima. The low temperature maximum is
centered around the optimal temperature, $T_c^{(\infty)}$,
corresponding to a partitioned strip, $v_c=\infty$. Most remarkably,
the optimal temperature of the second maximum on the right is
systematically higher than the optimal temperature,
$T_c^{(0)}$, of the un-partitioned strip, $v_c=0$. Moreover, such a double resonance could only be found for certain combinations of $v_c$ and $h/d$, as shown in Fig.\,\ref{F5}(b).

The occurrence of a double resonance can be explained by noticing
that for electrons with $v_x \leq v_c$, the wall acts as an
effective partition. In Fig.\,\ref{F5}(a), where $h=0$, this
corresponds to splitting the strip into two equal strips of half
width. Correspondingly, the slow electrons trapped in either half
strip contribute a cyclotron peak that is the highest for $T\simeq
T_c^{(\infty)}$. Fast electrons with $v_x>v_c$ are free to move
across the full width of the strip, so that the optimal temperature
of their cyclotron orbits ought to approach $T_c^{(0)}$ with
$T_c^{(0)} \simeq 4T_c^{(\infty)}$. However, by taking a closer look
at our derivation of Eq. (\ref{approx}), it is apparent that in the
case of fast electrons the lower limit of the integral must be
modified to account for the condition $v_x>v_c$. As a consequence of
the bell shaped profile of the Maxwell distribution, such a
modification of the integration range moves the optimal value $\bar
T_c$ to appreciably higher values, in agreement with Fig.\,\ref{F5}.
\section{Concluding remarks}
\label{conclusions}

The temperature controlled resonance of the cyclotron spectra,
emerging  from a matching between the system size length $d$ with
the thermal electron gyroradius $\lambda$, becomes detectable for a
confined magnetized gas of charged particles under two important
conditions, summarized by the inequalities $\gamma \ll \omega_c \ll
\omega_p$. The condition $\omega_c \ll \omega_p$,
introduced in Sec. \ref{infinite}, required applying magnetic fields of relatively low intensity. The underdamped regime, $\gamma \ll \omega_c$, was assumed to enhance the cyclotron peak of the transverse p.s.d., $S_x(\omega)$, over its background. Both conditions can be met in magnetoplasmas \cite{gentle_plasma}.

In normal metals the observation of the resonant cyclotron effects
reported here might seem out of question. At room temperature typical values
of the electron damping constant are $\gamma \sim 10^{13}$ s$^{-1}$,
or larger, so that an underdamped electron dynamics would set on
only for exceedingly large magnetic field \cite{Abri}. A more promising playground for an experimental demonstration of the effect under investigation is a 2D electron gas, where mobility can be quite high, thus, corresponding to a small damping constant, $\gamma \sim 10^{9}$
s$^{-1}$. A relatively low magnetic fields (of about $0.1$ T) then
would easily satisfy the condition $\gamma\ll \omega_c$. However,
since the plasma frequency, $\omega_p$, of an unconstrained 2D electron gas tend to be very low, artificial geometries should be implemented. To this regard it helps mentioning two sets of recent experiments, which detected, respectively, oscillations in the magnetoresistances of two-dimensional lateral surface superlattices with square patterns \cite{long}, and dynamical phase transitions between localized and superdiffusive (or ballistic) regimes for paramagnetic colloidal systems confined to magnetic bubble domains \cite{tierno}. Both results can be explained, in semiclassical approximation, as a commensurability effect between the cyclotron radius of the magnetocharges and the spatial periodicity of the substrate, without the need to invoke quantum mechanics.

We finally point out that the diffusion of confined magneto-charged particles is a topic of increasing interest not only in solid state physics. In medical research, for instance, magnetic nanostructures confined to 2D geometries are thought to offer the most exciting avenues to nanobiomagnetic applications, including targeted drug delivery, bioseparation and cancer therapy, even if their diffusion properties are not yet fully controllable. The possibility of extending our analysis to nanobiomagnetic processes at the cellular level requires advances on at least two issues: (1) diffusion of complex magnetic materials. In biomedical applications, pointlike magneto-charges are often replaced by synthetic magnetic structures, such as magnetic microdiscs with a spin-vortex ground state \cite{medicine}; (2) walls interactions. Contrary to our simple model, the interactions between a magneto-charge and cellular walls are typically inelastic, namely characterized by finite interaction times, energy transfer and even structural changes, like the activation on mechanosensitive ion channels. Both issues are the subject of ongoing investigations by research teams worldwide.

\acknowledgments FM acknowledges partial support from the Seventh Framework Programme under grant agreement n° 256959, project NANOPOWER.

\end{document}